\documentclass[a4paper]{jpconf}
\usepackage{graphicx}
\begin{document}
\title{Up- and down-quark contributions to the nucleon electromagnetic form factors at low $Q^2$}

\author{I A Qattan$^1$ and J Arrington$^2$}

\address{$^1$ Khalifa University of Science, Technology and Research, Department of Applied Mathematics and Sciences, P.O. Box 127788, Abu Dhabi, United Arab Emirates}
\address{$^2$ Physics Division, Argonne National Laboratory, Argonne, Illinois, 60439, USA}
\ead{issam.qattan@kustar.ac.ae}

\begin{abstract}
The spatial distribution of charge and magnetization within the nucleon (proton and neutron) is encoded in the
elastic electromagnetic form factors $G_E^{(p,n)}$ and $G_M^{(p,n)}$. These form factors have been precisely measured utilizing elastic electron scattering, and the combination of proton and neutron form factors allows for the separation of the up- and down-quark contributions to the nucleon form factors. We expand on our original analyses and extract the up- and down-quark contributions to the nucleon electromagnetic form factors from worldwide data with an emphasis on precise new data covering the low-momentum region, which is sensitive to the large-scale structure of the nucleon. From these, we construct the flavor-separated Dirac and Pauli form factors and their ratios, and compare the results to recent extractions and theoretical calculations and models.
\end{abstract}

\section{Introduction}
The nucleon electromagnetic form factors $G_E^{(p,n)}(Q^2)$ and $G_M^{(p,n)}(Q^2)$ are fundamental quantities in
nuclear and elementary particle physics as they provide information on the spatial distributions of charge and magnetization within the nucleon. They are a function of the four-momentum transfer squared of the virtual photon, $Q^2$. In the nonrelativistic limit, they are the Fourier transform of the charge and magnetization distributions.  Therefore, isolating the up- and down-quark contributions to the nucleon form factors is essential to examine spatial asymmetries in the quark distributions.

There are primarily two methods used to extract the proton form factors. The first is the Rosenbluth or longitudinal-transverse (LT) separation method~\cite{rosen50} which uses measurements of unpolarized cross section, and the second is the polarization transfer/polarized target (PT) method~\cite{dombey69} which requires measurements of the spin-dependent cross section.  In the one-photon exchange (OPE) approximation or the Born value, the unpolarized cross section is proportional to the "reduced" cross section, $\sigma_{R}= G_M^2+(\varepsilon/\tau) G_E^{2}$, where $\tau=Q^2/4M_{N}^2$, $M_N$ is the nucleon mass, and $\varepsilon$ is the virtual photon longitudinal polarization parameter defined as $\varepsilon^{-1} = [1+2(1+\tau)\mbox{tan}^{2}(\theta_{e}/2)]$, where $\theta_{e}$ is the scattering angle of the electron.  Measuring $\sigma_{R}$ at several $\varepsilon$ points for a fixed $Q^2$ value, one can separate $G_E^p$ and $G_M^p$.  However, for cases where $\varepsilon/\tau$ is extremely small (large), it is difficult to extract $G_E^p$ ($G_M^p$) with precision. On the other hand, the polarization measurements are sensitive only to the ratio $G_E^p/G_M^p$. Therefore, by taking ratios of polarization components, many of the systematic uncertainties in the polarization measurements cancel, allowing for precise measurements of the ratio $R_p = \mu_{p} G_E^p/G_M^p$~\cite{puckett2012}, where $\mu_{p}$ is the proton magnetic dipole moment. The two methods yield strikingly different results for the ratio $R_p$ in the region $Q^2 > 2$~(GeV/c)$^2$~\cite{arrington03}, where the Rosenbluth extractions show approximate scaling, $\mu_{p} G_E^p/G_M^p \approx 1$, while the recoil polarization data indicate a nearly linear decrease in $R_p$ with $Q^2$ with some hint of flattening out above 5 (GeV/c)$^2$. Recent studies suggest that hard two-photon exchange (TPE) corrections to the unpolarized cross section may resolve the discrepancy~\cite{arrington07,CV2007,ABM2011,qattan2011a,adikaram2015}.

\section{Extraction of the flavor-separated form factors}
Recent precise measurements of the neutron's electric to magnetic form factor ratio $R_n = \mu_{n} G_E^n/G_M^n$ up to 3.4~GeV$^2$~\cite{riordan2010}, combined with existing $R_p = \mu_{p} G_E^p/G_M^p$ measurements in the same
$Q^2$ range allowed for a separation of the up- and down-quark contributions to the nucleon form factors
at large $Q^2$~\cite{CJRW2011}. This first analysis, referred to as ``CJRW'' in this work, examined the scaling behavior of the up- and down-quark contributions at large $Q^2$, supporting the idea that diquark correlations play an important role~\cite{cloet09}. Recently, we extended the flavor separation analysis~\cite{qattan2012,qattan2014,qattan2015} using combined cross section and polarization measurements of elastic electron-proton scattering with an emphasis on precise new data from Ref.~\cite{bernauer2014} covering the low-momentum region, which is sensitive to the large-scale structure of the nucleon. In our work, we account for effects neglected in the original work where we apply TPE corrections in the extraction of the proton form factors based on the approach of Ref.~\cite{qattan2011b}. The TPE correction applied in our work, based on the parametrization from Ref.~\cite{borisyuk-kobushkin2011}, is linear in $\varepsilon$~\cite{tvaskis2006} and vanishes in the limit $\varepsilon \rightarrow$ 1~\cite{arrington11-chen07,arrington04}. We also compare our results to a parametrization of the proton form factors extracted~\cite{arrington07} after applying the hadronic calculation for TPE from Ref.~\cite{bmt2003}. We also include additional new $G_M^n$ data from CLAS~\cite{lachniet09} and performed a new global fit to $G_M^n$ which we used, along with the parametrization of $R_n$~\cite{riordan2010}, to construct $G_E^n$, as well as account for the uncertainties associated with all of the form factors measurements as the CJRW analysis accounted only for uncertainties on $R_n$ which was the dominant uncertainty for their flavor-separated results. Finally, we use our results of the flavor-separated form factors to construct the flavor-separated Dirac, $F_1^{(u,d)}$, and Pauli, $F_2^{(u,d)}$, form factors and their ratios.

\section{Results and discussion}
In this section, we present our results of the flavor-separated form factors $F_1^{(u,d)}$ and $F_2^{(u,d)}$. We then compare our results to the CJRW extractions which allows for examination of the effect of the TPE corrections applied, additional uncertainties, as well as updated form factor data set used. We also compare our results to the Venkat {\textit{et al.}}~\cite{venkat2011} ("VAMZ"), and Arrington {\textit{et al.}}~\cite{arrington07} ("AMT") extractions which use improved proton form factor parametrization obtained assuming different treatment of TPE corrections at lower $Q^2$ values. In addition, we used the Venkat plus the $G_M^n$ and
$R_n$ fits mentioned above, and looked at the impact of our updated $G_M^n$ fit by replacing this with the Kelly~\cite{kelly2004} fit for $G_M^n$ ("VAMZ-Kelly"). Finally, we compare the results to recent theoretical calculations and fits to the flavor-separated form factors: a Dyson-Schwinger equation ("DSE") calculation~\cite{cloet09}, a pion-cloud relativistic constituent quark model ("PC-RCQM")~\cite{cloet2012}, a relativistic constituent quark model whose hyperfine interaction is derived from Goldstone-boson exchange ("GBE-RCQM")~\cite{rohrmoser2011}, and a generalized parton distribution (GPD) calculations~\cite{gonzalez2013}.

%%%% Flavors Structure Functions:
\begin{figure}
\includegraphics[width=5.29cm,clip,bb=28 180 525 603]{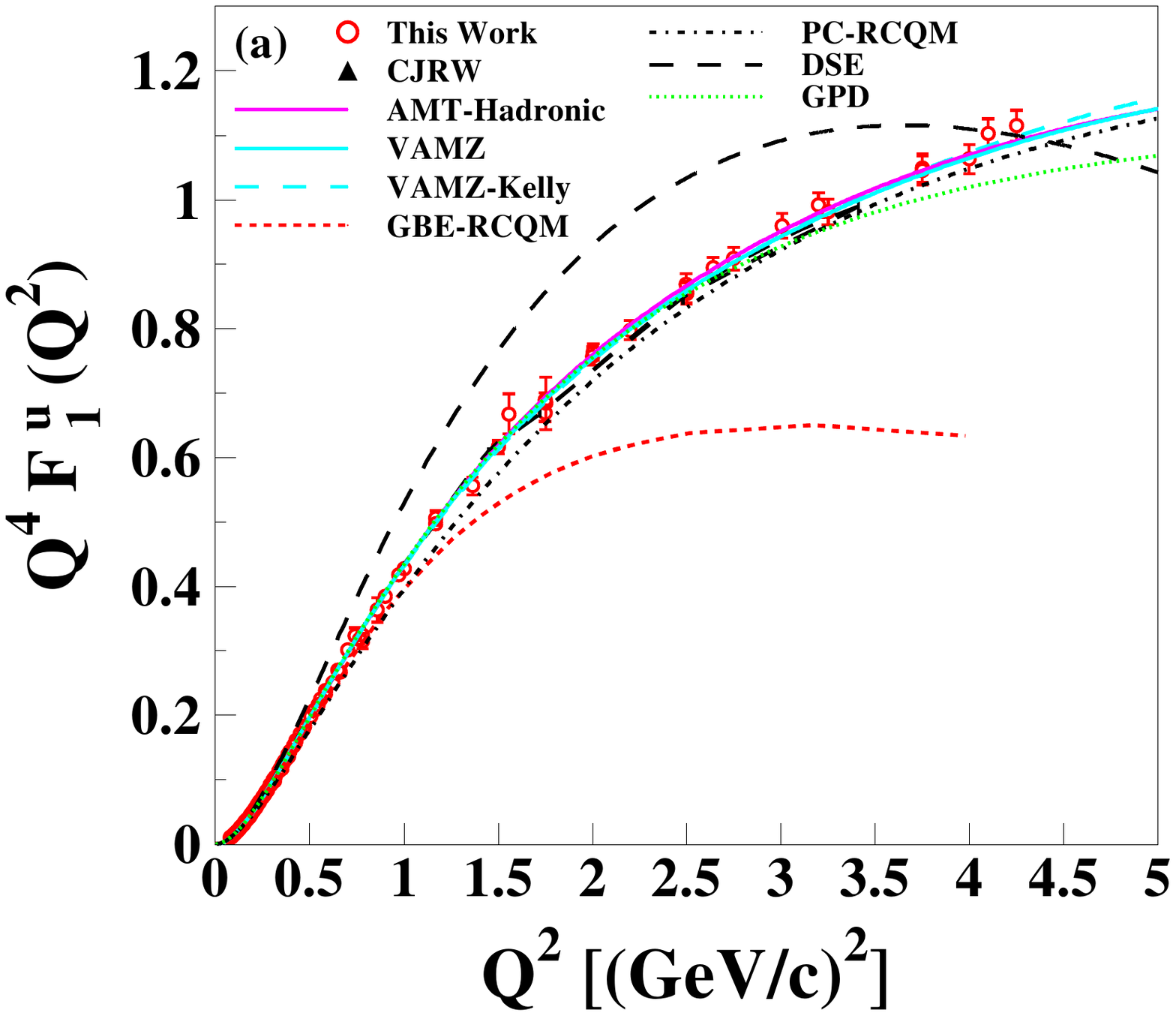}
\includegraphics[width=5.29cm,clip,bb=21 180 525 603]{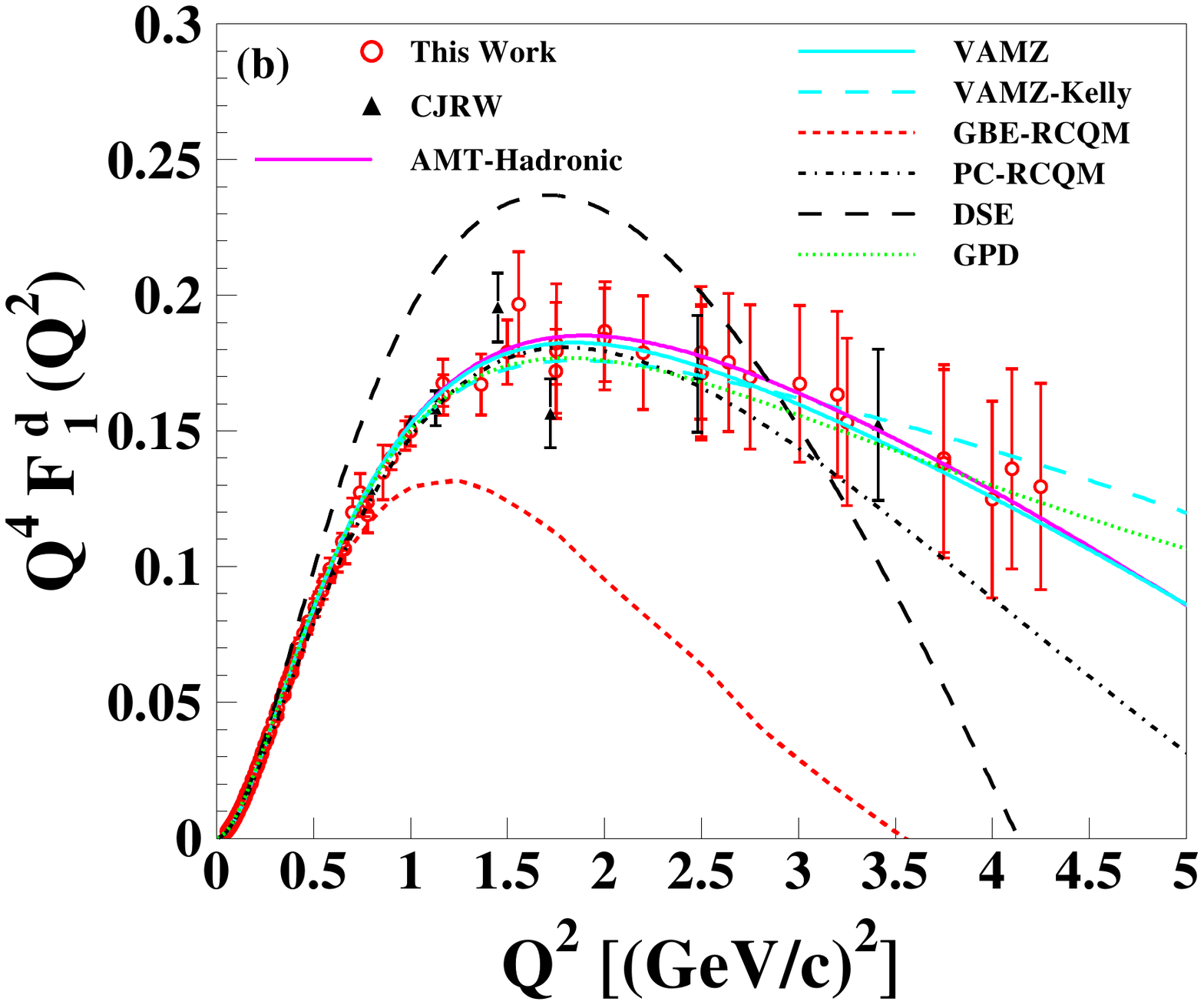}
\includegraphics[width=5.29cm,clip,bb=21 180 525 603]{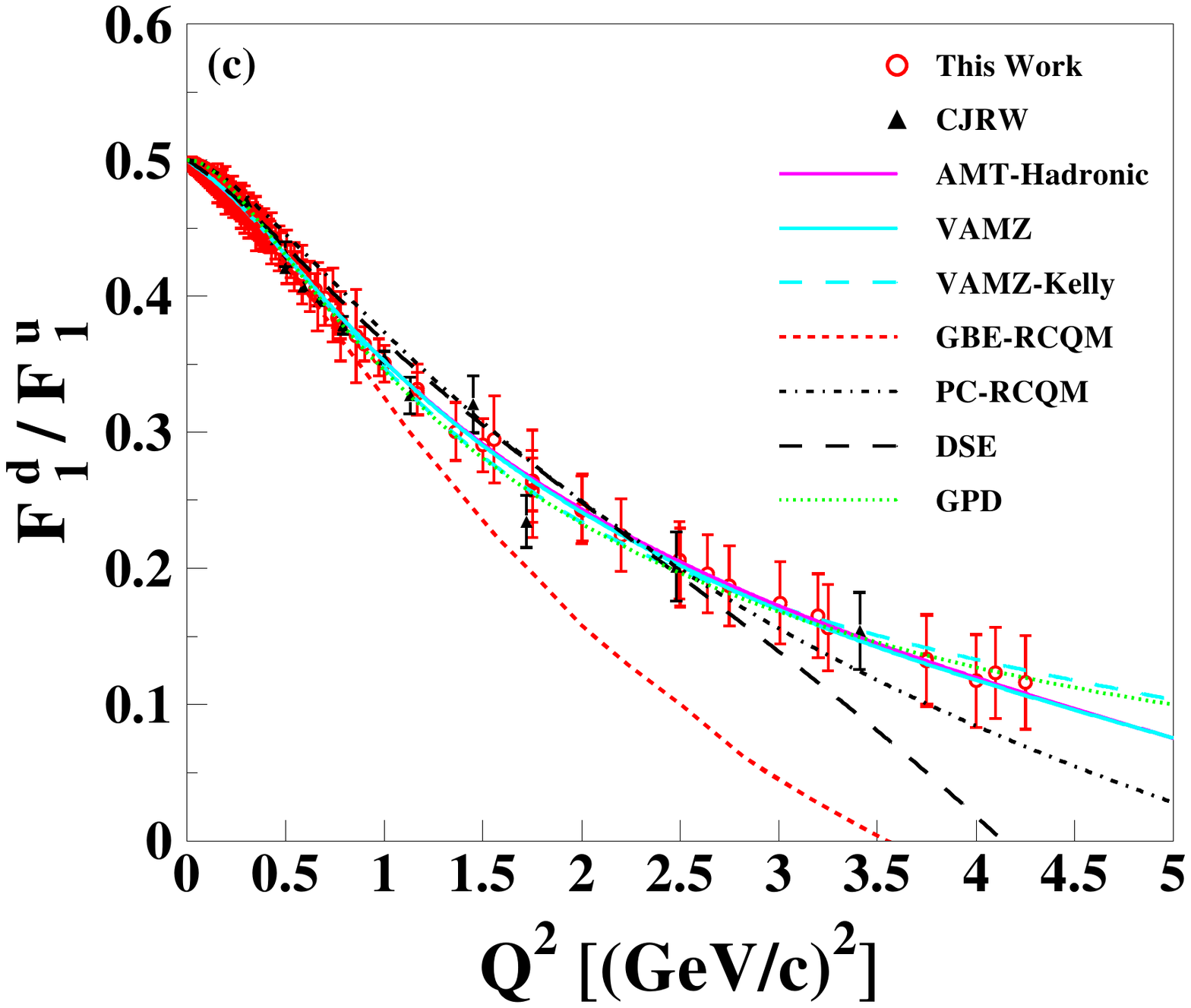}\\
\includegraphics[width=5.29cm,clip,bb=21 180 525 603]{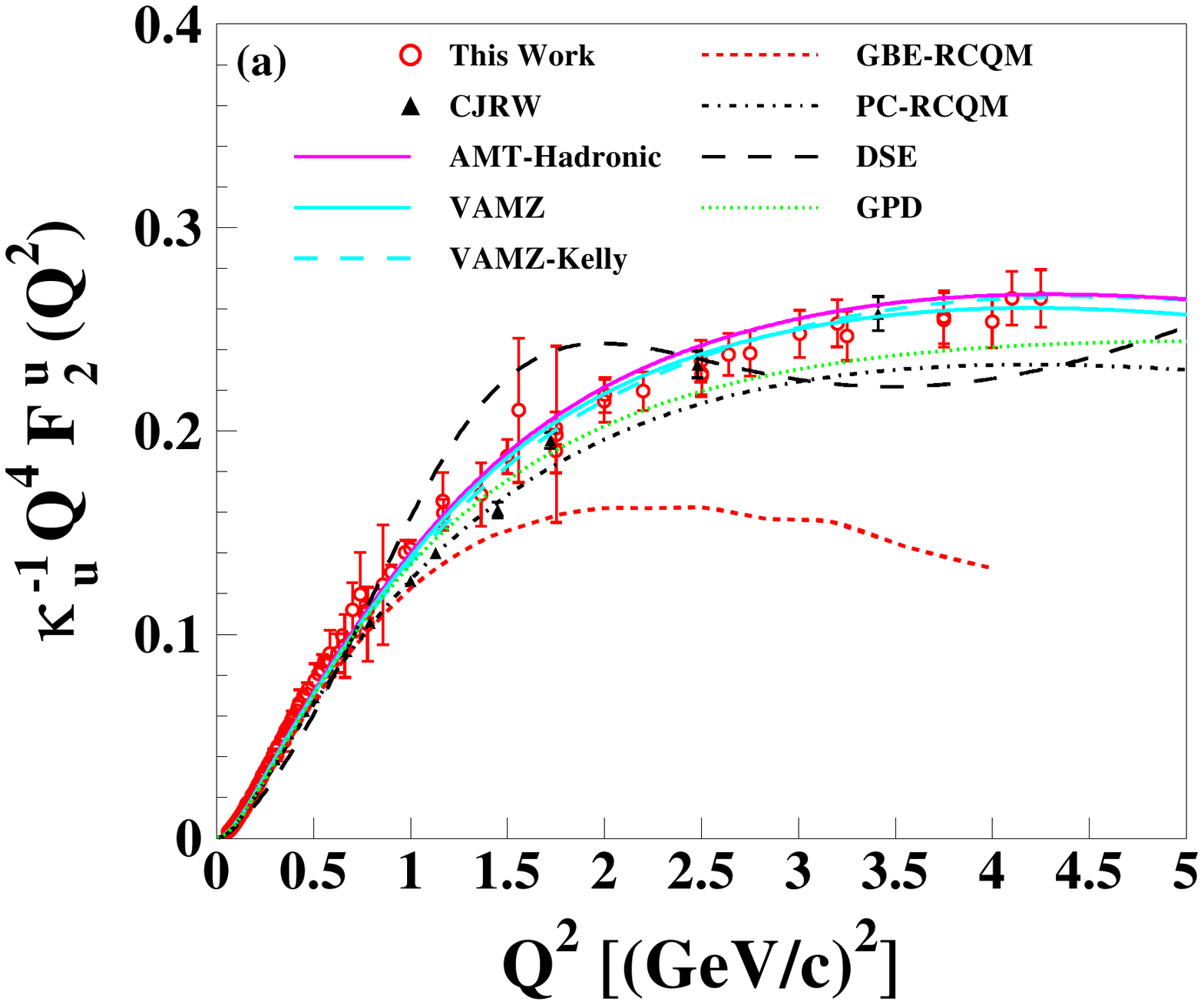}
\includegraphics[width=5.29cm,clip,bb=21 180 525 603]{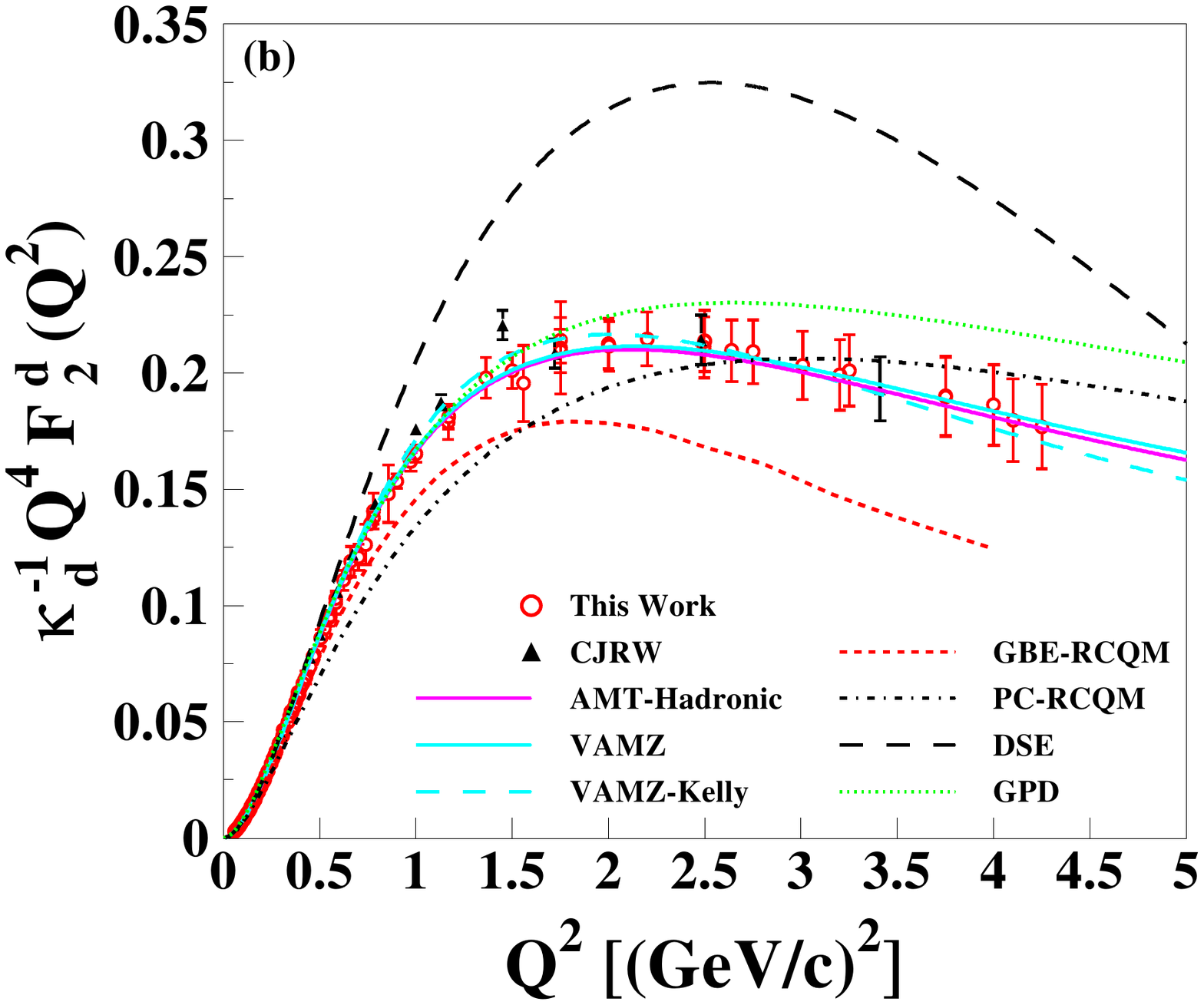}
\includegraphics[width=5.29cm,clip,bb=21 180 525 603]{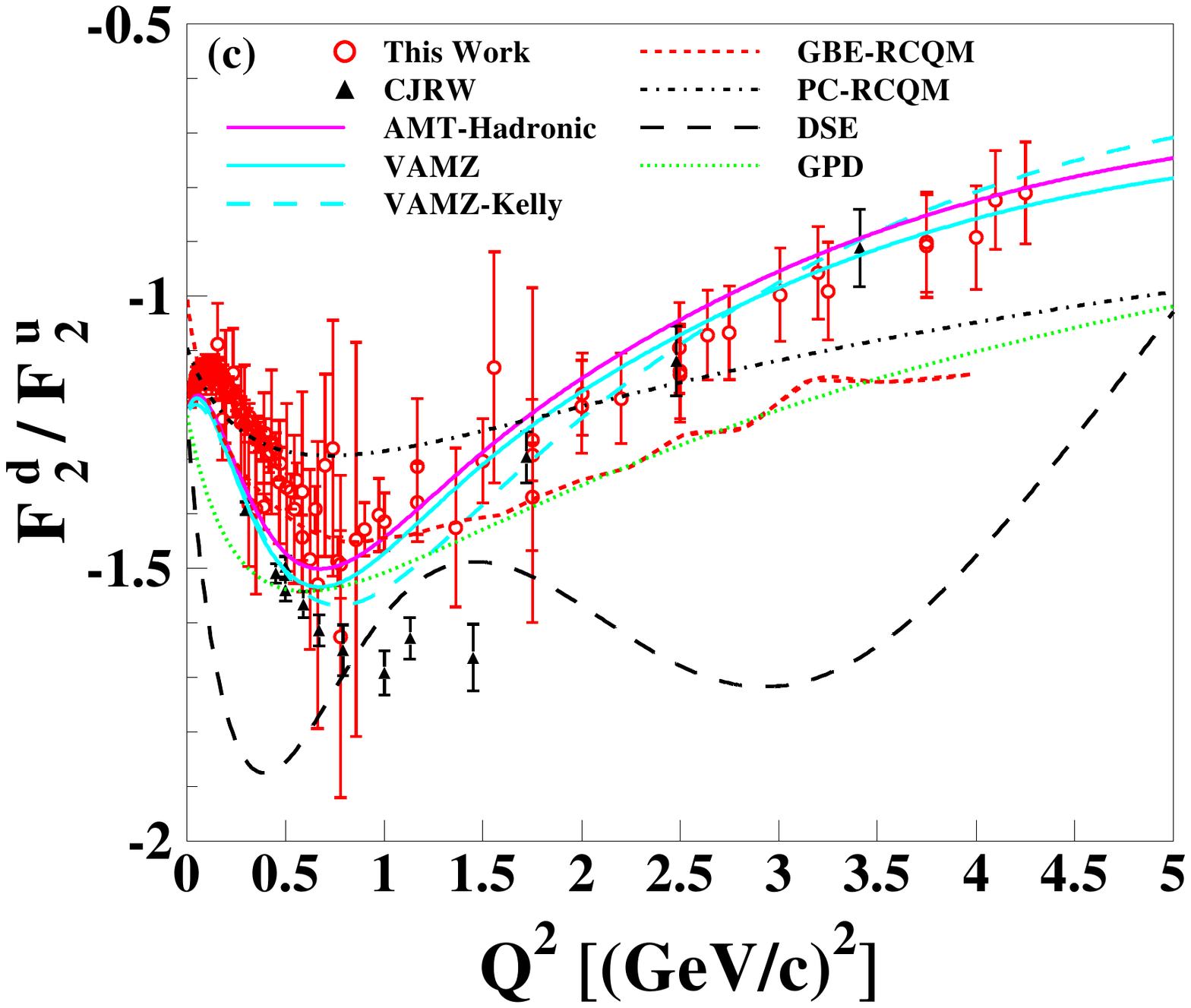}
\caption{The flavor-separated form factors $F_1^{(u,d)}$ (top) and $F_2^{(u,d)}$ (bottom) and their ratios from our analysis~\cite{qattan2015} and the CJRW extractions~\cite{CJRW2011}. Also shown are the AMT~\cite{arrington07} and VAMZ fits~\cite{venkat2011}, and the values from the GBE-RCQM~\cite{rohrmoser2011}, PC-RCQM~\cite{cloet2012}, the DSE~\cite{cloet09}, and the GPD~\cite{gonzalez2013} models.}
\label{fig1}
\end{figure}

Figure~\ref{fig1} shows the flavor-separated contributions of $F_1$ and $F_2$. It was reported in the CJRW analysis that while both the up-quark form factors, $F_1^u$ and $F_2^u$, continued to rise compared to the down-quark, the down-quark contributions, $F_1^d$ and $F_2^d$, strikingly exhibited $1/Q^4$ scaling behavior above $Q^2 =$ 1.0 (GeV/c)$^2$ in agreement with the moments of the generalized parton distributions predictions~\cite{diehl2005}. However, these predictions are based on fits which included nucleon form factors data, except for the most recent $G_E^n$ data. Our results for $F_1^u$ and $F_2^u$ suggest as well that both form factors continue to rise with increasing $Q^2$ but with $F_2^u$ values somewhat higher than the CJRW extractions at low $Q^2$. This can be seen clearly in the ratio $F_2^d /F_2^u$. On the other hand, our results suggest that the down-quark contributions are falling more rapidly than the up-quark contributions at high $Q^2$ with respect to the $1/Q^4$ behavior which is more apparent for $F_2^d$ and the ratio $F_2^d /F_2^u$ in agreement with global parametrizations and theoretical calculations which predict faster falloff than the apparent $1/Q^4$ scaling behavior. However, such behavior is sensitive
to the parametrization of $G_E^n$ at $Q^2 >$ 2 (GeV/c)$^2$~\cite{riordan2010}. The difference in the results obtained are attributed mainly to the TPE corrections applied to the proton form factors and to lesser extent to the use of the updated $G_M^n$ parametrization. In addition, while the $G_E^n$ uncertainties have the largest impact, the additional uncertainties from the proton and $G_M^n$ yield a non-negligible increase in the total uncertainties.

%%%% Flavors Structure Functions Ratios:
\begin{figure}
\hspace{0.5cm}
\includegraphics[width=7.0cm,clip,bb=20 180 525 603]{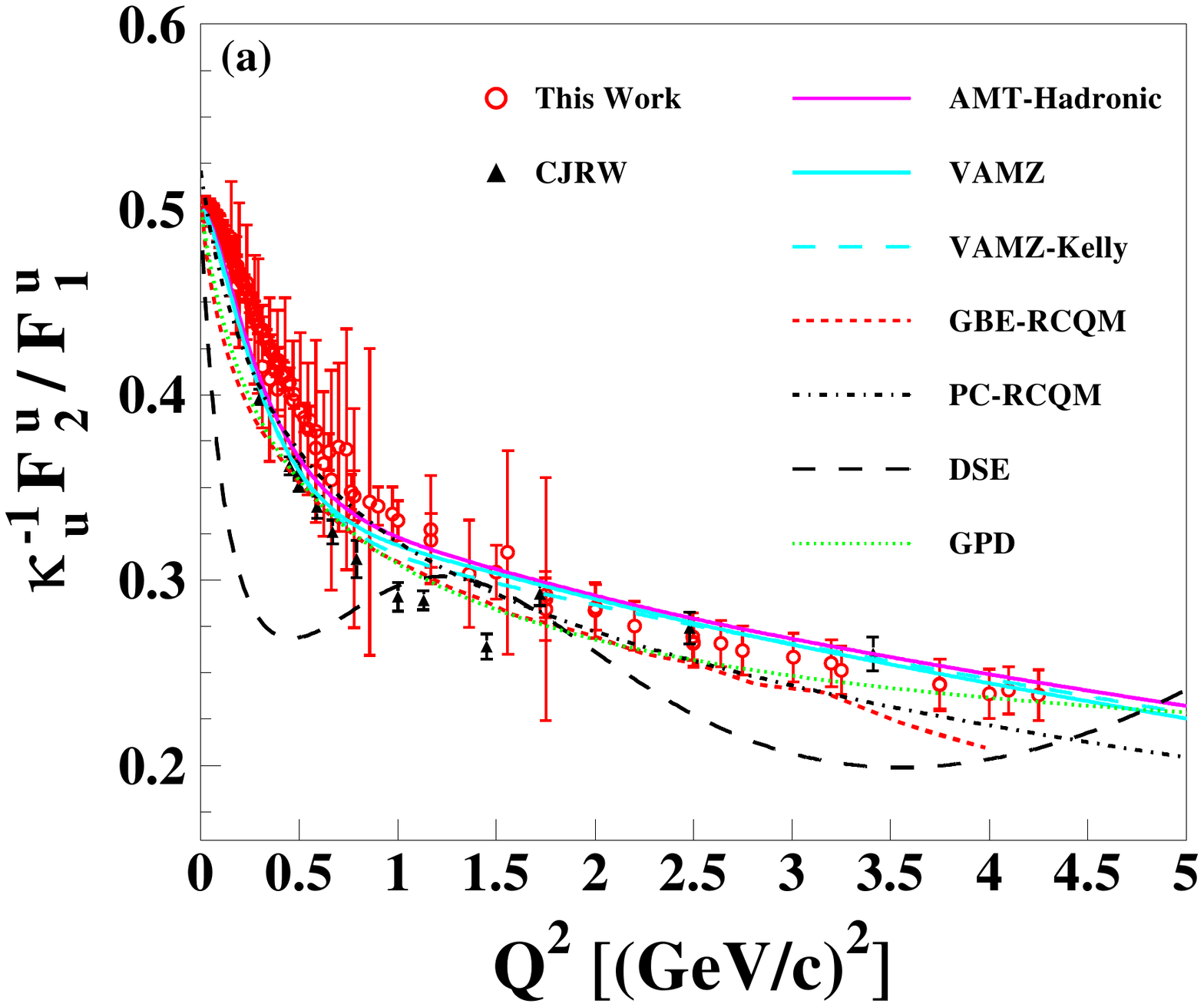} \hspace{0.5cm}
\includegraphics[width=7.0cm,clip,bb=20 180 525 603]{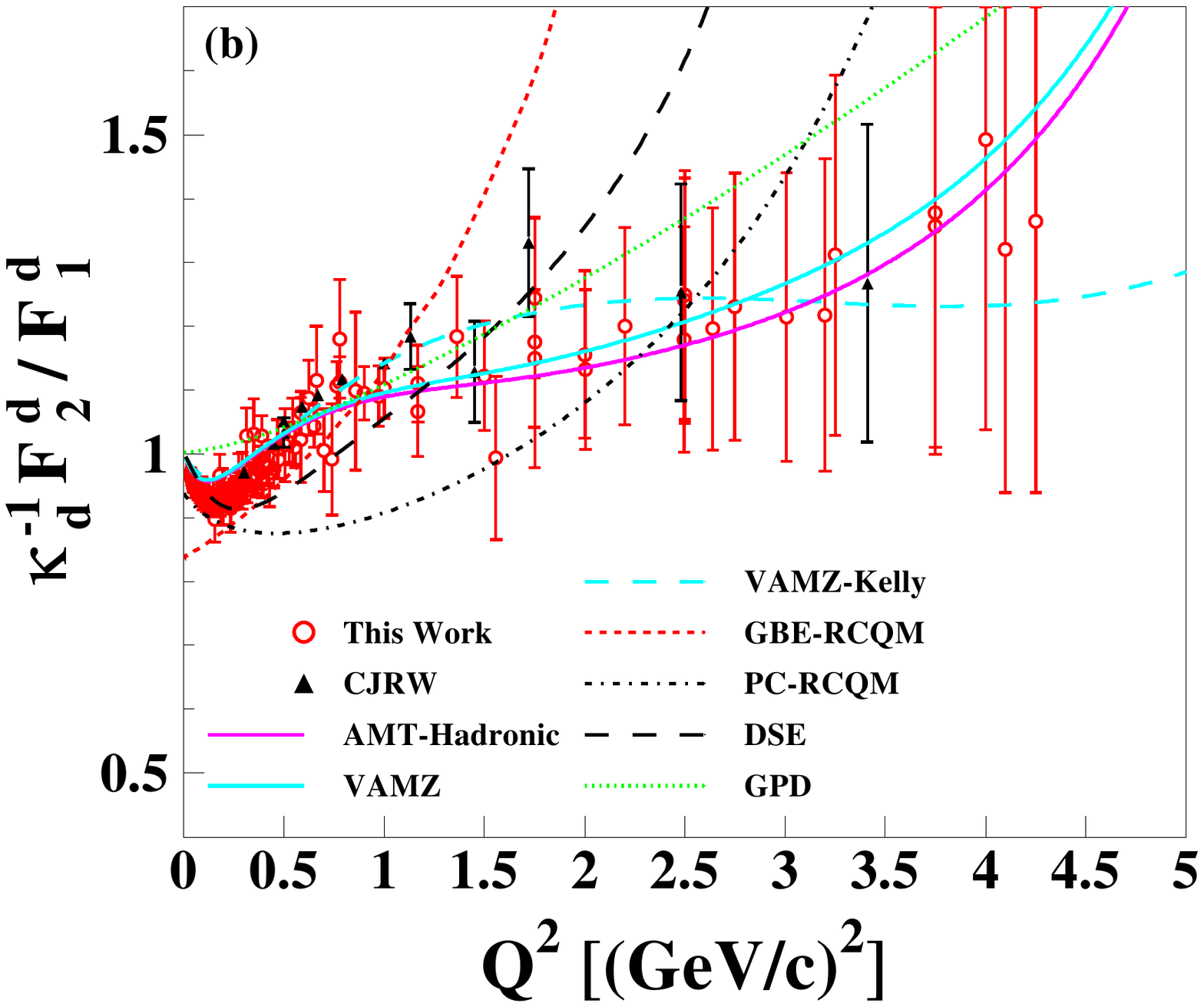}\\
\vspace{-0.5cm}
\caption{The ratios $\kappa_u^{-1} F_2^u /F_1^u$ and $\kappa_d^{-1} F_2^d /F_1^d$. Points and curves as in Fig.~\ref{fig1}.}
\label{fig2}
\end{figure}

Figure~\ref{fig2} shows the ratios $\kappa_u^{-1} F_2^u /F_1^u$ and $\kappa_d^{-1} F_2^d /F_1^d$. Here  $\kappa_{u,d}$ are the $Q^2 =$ 0 limits of $F_2^{u,d}$ with $\kappa_u = \mu_u -2 =$ 1.67 and $\kappa_d = \mu_d - 1 =$ $-$ 2.03. The ratios are scaled by $\kappa_{u,d}^{-1}$ and normalized to $1/F_1^{u,d}$ yielding 0.5(1) for the up(down)-quark contribution. At low $Q^2$, the ratio $F_2^u /F_1^u$ falls rapidly up to $Q^2 \simeq 1$ (GeV/c)$^2$ where it starts to decrease slowly. While our values are somewhat larger than those obtained in the CJRW extractions for $Q^2 <$ 1.5 (GeV/c)$^2$ due to the difference in the $F_2^u$ values, they are in good qualitative agreement with form factor fits and calculations. The ratio $\kappa_d^{-1} F_2^d /F_1^d$ increases slowly at low $Q^2$ and becomes almost constant for $Q^2 >$ 1 (GeV/c)$2$. Our results are consistent with both the CJRW results and extractions based on form factors fits. However, our results differ from all calculations as they predict a rapid falloff of $F_1^d$ at large $Q^2$ compared to $F_2^d$ which clearly leads to a rapid rise in the ratio. The high-$Q^2$ measurements of $G_E^n$ planned at Jefferson Lab after the 12 GeV upgrade~\cite{dudek2012} are clearly critical to pin down the behavior of this ratio as well as to examine the theoretical calculations and models.

\ack{}
This work was supported by Khalifa University of Science, Technology and Research and by the U.S. Department of
Energy, Office of Nuclear Physics, under contract DE-AC02-06CH11357.


\section*{References}
\begin{thebibliography}{30}
%%%(1)
\bibitem{rosen50} Rosenbluth M N 1950 {\it Phys. Rev.} {\textbf{79}} 615
%%%(2)
\bibitem{dombey69} Dombey N 1969 {\it Rev. Mod. Phys.} {\textbf{41}} 236
%%%(3)
\bibitem{puckett2012}
Punjabi V {\textit{et al.}} 2005 {\it Phys. Rev.} \textbf{C71} 055202; Puckett A J R {\textit{et al.}} 2012 {\it Phys. Rev.}  \textbf{C85} 045203; Puckett A J R {\textit{et al.}} 2010 {\it Phys. Rev. Lett.} \textbf{104} 242301; Zhan X \textit{et al.} 2011 {\it Phys. Lett.} \textbf{B705} 59; Ron G \textit{et al.} 2011 {\it Phys. Rev.} \textbf{C84} 055204; Arrington J, Perdrisat C, and de Jager C W 2011 {\it J. Phys. Conf. Ser.} \textbf{299} 012002
%%%(4)
\bibitem{arrington03}
Qattan I A {\textit{et al.}} 2005 {\it Phys. Rev. Lett.} \textbf{94} 142301; Arrington J 2003 {\it Phys. Rev.} \textbf{C68} 034325
%%%(5)
\bibitem{arrington07}
Arrington J, Melnitchouk W, and Tjon J A 2007 {\it Phys. Rev.} \textbf{C76} 035205
%%%(6)
\bibitem{CV2007}
Carlson C E and Vanderhaeghen M 2007 {\it Ann. Rev. Nucl. Part. Sci.} \textbf{57} 171
%%%(7)
\bibitem{ABM2011}
Arrington J, Blunden P, and Melnitchouk W 2011 {\it Prog. Part. Nucl. Phys.} \textbf{66} 782
%%%(8)
\bibitem{qattan2011a}
Qattan I A and Alsaad A 2011 {\it Phys. Rev.}  \textbf{C84} 054317 [ Erratum: 2011 {\it Phys. Rev.} \textbf{C84} 029905]
%%%(9)
\bibitem{adikaram2015}
Adikaram D {\textit{et al.}} (CLAS Collaboration) 2015 {\it Phys. Rev. Lett.} \textbf{114} 062003; Rachek I A {\textit{et al.}} 2015 {\it Phys. Rev. Lett.} \textbf{114} 062005; Rimal D {\textit{et al.}} (CLAS Collaboration) 2016 {\it Preprint} arXiv:1603.00315 [nucl-ex]
%%%(10)
\bibitem{bmt2003}
Blunden P G, Melnitchouk W, and Tjon J A 2005 {\it Phys. Rev.} \textbf{C72} 034612
%%%(11)
\bibitem{riordan2010}
Riordan S {\textit{et al.}} 2010 {\it Phys. Rev. Lett.} \textbf{105} 262302
%%%(12)
\bibitem{CJRW2011}
Cates G, de Jager C, Riordan S, and Wojtsekhowski B 2011 {\it Phys. Rev. Lett.} \textbf{106} 252003
%%(13)
\bibitem{cloet09}
Cloet I {\textit{et al.}} 2009 {\it Few-Body Syst.} \textbf{46} 1
%%%(14)
\bibitem{qattan2012}
Qattan I A and Arrington J 2012 {\it Phys. Rev.} \textbf{C86} 065210
%%%(15)
\bibitem{qattan2014}
Qattan I A and Arrington J 2014 {\it Eur. Phys. J. WOC} \textbf{66} 06020
%%%(16)
\bibitem{qattan2015}
Qattan I A, Arrington J, and Alsaad A 2015 {\it Phys. Rev.} \textbf{C91} 065203
%%%(17)
\bibitem{bernauer2014}
Bernauer J C {\textit{et al.}} (A1 Collaboration) 2014 {\it Phys. Rev.} \textbf{C90} 015206
%%%(18)
\bibitem{qattan2011b}
Qattan I A, Alsaad A, and Arrington J 2011 {\it Phys. Rev.} \textbf{C84} 054317
%%%(19)
\bibitem{borisyuk-kobushkin2011}
Borisyuk D and Kobushkin A 2011 {\it Phys. Rev.} \textbf{D83} 057501
%%%(20)
\bibitem{tvaskis2006}
Tvaskis V {\textit{et al.}} 2006 {\it Phys. Rev.} \textbf{C73} 025206
%%%(21)
\bibitem{arrington11-chen07}
Arrington J, Blunden P G, and Melnitchouk W 2011 {\it Prog. Part. Nucl. Phys.} \textbf{66} 782;
Chen Y -C, Kao C -W, and Yang S -N 2007 {\it Phys. Lett.} \textbf{B652} 269
%%%(22)
\bibitem{arrington04}
Arrington J 2004 {\it Phys. Rev.} \textbf{C69} 032201
%%%(23)
\bibitem{lachniet09}
Lachniet J {\textit{et al.}} (CLAS Collaboration) 2009 {\it Phys. Rev. Lett.} \textbf{102} 192001
%%%(24)
\bibitem{venkat2011}
Venkat S, Arrington J, Miller G A, and Zhan X 2011 {\it Phys. Rev.} \textbf{C83} 015203
%%%(25)
\bibitem{kelly2004}
Kelly J J 2004 {\it Phys. Rev.} \textbf{C70} 068202
%%%(26)
\bibitem{cloet2012}
Cloet I C and Miller G A 2012 {\it Phys. Rev.} \textbf{C86} 015208
%%%(27)
\bibitem{rohrmoser2011}
Rohrmoser M, Choi K -S, and Plessas W 2011 {\it Preprint} arXiv:1110.3665 [hep-ph]
%%%(28)
\bibitem{gonzalez2013}
Gonzalez-Hernandez J O, Liuti S, Goldstein G R, and Kathuria K 2013 {\it Phys. Rev.} \textbf{C88} 065206
%%%(29)
\bibitem{diehl2005}
Diehl M, Feldmann T, Jakob R, and Kroll P 2005 {\it Eur. Phys. J.} \textbf{C39} 1
%%%(30)
\bibitem{dudek2012}
Dudek J {\textit{et al.}} 2012 {\it Eur. Phys. J.} \textbf{A48} 187
\end{thebibliography}
\end{document}